\documentclass[review,onefignum,onetabnum]{siamart171218}

\usepackage{lipsum}
\usepackage{amsfonts}
\usepackage{graphicx}
\usepackage{epstopdf}
\usepackage{algorithmic}
\usepackage{comment}
\ifpdf
  \DeclareGraphicsExtensions{.eps,.pdf,.png,.jpg}
\else
  \DeclareGraphicsExtensions{.eps}
\fi

\newsiamremark{remark}{Remark}
\newsiamremark{hypothesis}{Hypothesis}
\crefname{hypothesis}{Hypothesis}{Hypotheses}
\newsiamthm{claim}{Claim}

\headers{American Options Pricing}{}

\title{An Alternative Approach to Evaluate American Options Price Using HJM Approach}

\author{Kushantha Fernando\thanks{Independent; *Corresponding author: kushfernando@gmail.com }
\and Vajira Manathunga\thanks{Middle Tennessee State University, Murfreesboro, TN; Vajira.Manathunga@mtsu.edu}}

\usepackage{amsopn}

\makeatletter
\newcommand*{\addFileDependency}[1]{
  \typeout{(#1)}
  \@addtofilelist{#1}
  \IfFileExists{#1}{}{\typeout{No file #1.}}
}
\makeatother

\ifpdf
\hypersetup{
  pdftitle={An Alternative Approach to Evaluate American Options Price Using HJM Approach},
  pdfauthor={Kushantha Fernando, Vajira Manathunga}
}
\fi

\DeclareMathOperator*{\esssup}{ess\,sup}
\usepackage{cool}
\usepackage{bbm}
\usepackage{verbatim}
\usepackage{natbib}
\begin{document}

\maketitle

\begin{abstract}
Developments in finance industry and academic research has led to innovative financial products. This paper presents an alternative approach to price American options. Our approach utilizes famous \cite{heath1992bond} (``HJM") technique to calculate American option written on an asset. Originally, HJM forward modeling approach was introduced as an alternative approach to bond pricing in fixed income market. Since then,  \cite{schweizer2008term} and  \cite{carmona2008infinite} extended HJM forward modeling approach to equity market by capturing dynamic nature of volatility. They modeled the term structure of volatility, which is commonly observed in the market place as opposed to constant volatility assumption under Black - Scholes framework. 
Using this approach, we propose an alternative value function, a stopping criteria and a stopping time. We give an example of how to price American put option using proposed methodology.

\end{abstract}

\begin{keywords}
HJM Model, American option pricing, Arbitrage-free
\end{keywords}

\begin{AMS}
  91B05, 91G15, 91G80
\end{AMS}

\section{Introduction}
Financial markets have gotten increasingly complex over the past few decades due to increasing demand for various financial products. These
instruments varies from basic vanilla derivatives such as Forwards, Futures, Swaps to complex
and exotic derivative contracts such as look back options to Asian, American to barrier type
options. Some of these exotic products may not even have a closed form theoretical solution
and arbitrage free valuations of these products be quite complex.\\ 

This paper seek to introduce an alternative pricing mechanism of one such derivative
contract: An American option. The American option is a financial contract written on another asset
with option buyer has the right to exercise the option at any given time during it’s lifetime.
Theoretical pricing of American option and it’s implementation via numerical schemes, Monte Carlo simulation methods and
various lattice schemes has been already studied and there are numerous publications
currently exist in financial literature. In this paper we propose an alternative approach to price American option using \cite{heath1992bond}(HJM) forward modeling methodology.\\

HJM method was originally introduced in 1992 in fixed income market as an alternative method
for bond pricing based on forward interest rate compared to bond pricing based on short rate.
Under forward modeling philosophy, Heath-Jarrow-Morton propose a method to price bonds
via modeling the evolution of the entire yield curve while taking initial yield curve as a model
input. Other than simply providing an arbitrage free evolution of forward curve, as a
consequence, HJM approach also provides a way to valuation of derivative contracts using forward evolution of yield curves. There has been many applications of HJM approach appeared in academia as well as in industry since then. In recent years,this forward modeling
paradigm was extended to equity market as well to energy market \cite{broszkiewicz2006calibration}, interest futures \cite{kamizono1996implementation}, credit and equity market \cite{carmona2007hjm}.\\

First extension of HJM philosophy to equity market can be found in 
\cite{schweizer2008term} and  \cite{carmona2007hjm}. They employed forward modeling philosophy by
dynamic modeling of forward volatility curves to facilitate European type option pricing. Such
forward evolution of volatility curves helps to relax the constant volatility assumption of Black –Scholes model and be more practical and realistic about under Black-Scholes framework. 

As stated earlier, we present an alternative approach to solve traditional American option
problem utilizing of HJM forward modeling approach. Since American option valuation is a
essentially an optimal stopping problem, it’s value is given by Snell envelop of the value process. By adapting HJM method discussed \cite{carmona2007hjm}, we reformulate a new value process for American option based on drift of the underlying asset. We claim that value of the American option depends on the forward drift of the underlying asset. In addition to new drift-based value function for the American type option written on an asset, we provide corresponding stopping criteria and stopping time
associated with the optimal stopping problem. Furthermore, related spot consistency conditions for
the forward drift is examined to be consistent with HJM modeling philosophy. We propose two
valuation methods called additive model and multiplicative model under this proposed
forward modeling philosophy. Then we use American put option pricing to illustrate our model under additive model under Black – Scholes assumptions.

\section{Additive Model}
\label{sec:main}
Let $(\Omega,\mathcal{F},P)$ be a probability space equipped with complete and right continuous filtration $(\mathcal{F}_t)_{t\ge 0}$, and $(G_t)_{t\ge 0}$ be a stochastic process adapted to the filtration $(\mathcal{F}_t)_{t\ge 0}$. We assume that the filtration is the Brownian filtration.  An American option is a financial derivative which allows the option holder to exercise the option at any time during its lifetime. Assume that risk free rate is constant and denoted by $r>0$, underlying stock pays no dividend and American options follows the ``gain"(payoff) process $(G_t)_{t\ge 0}$. 
The value of an American option at time $t$ is given by \cite{peskir2006optimal}
\begin{equation}
    V_t(T)=\sup_{t\le \tau\le T}E[e^{-r(\tau-t)}G_{\tau}|\mathcal{F}_t]
\end{equation} where $\tau$ is a stopping time and $T$ is the expiration time. Assume $T<\infty$, at every stopping time $\tau\in[0,T]$, the holder must decide whether or not to exercise the option. Thus this is a finite time horizon optimal stopping problem. In the event, where underlying stock is paying dividend continuously at a rate of $\delta$, replace $r$ by $(r-\delta)$  \\

Now assume that the gain process $G=(G_t)_{t\ge 0}$ is right and left continuous over stopping time $\tau$, $E[\sup_{0\le \tau \le T}e^{-(r)(\tau-t)}|G_{\tau}|]<\infty$. Then the solution to the above defined optimal stopping problem is given by the Snell envelop or essential supremum of $G$ defined as follows \cite{peskir2006optimal}.
\begin{equation}
    V_t(T)=\esssup_{t\le\tau\le T} E[e^{-r(\tau-t)}G_{\tau} | \mathcal{F}_t]
\end{equation}  with the optimal stopping time given by 
\begin{equation}\label{eq3}
\tau^*=\tau_t=\inf\{\,t\le s\le T| V_s=G_s\}\,
\end{equation}
It is worthwhile to observe that, if the market is free of arbitrage, then $\{\,e^{-rt}V_t\}\,_{t\ge 0}=\{\,E[e^{-r\tau}G_{\tau}|\mathcal{F}_t]\}\,$ are $P$-martingales \cite{carmona2007hjm}.

\subsection{HJM approach to American option pricing under additive model} 
Assume that the dynamics of the gain process $G_t$ is given by
\begin{equation}\label{eq1}
dG_t=\mu_tdt+\sigma_tdW_t 
\end{equation}
where $\int_{t}^{T}|\mu_u|du<\infty$ P-a.s and $\int_{t}^{T}|\sigma^2_u|du<\infty$ P-a.s.Since $\tau^*$ is optimal stopping time, we have 
\begin{equation}\label{eqvalue}
V_t(T)=E[e^{-r(\tau^*-t)}G_{\tau^*} | \mathcal{F}_t]
\end{equation}
Following the approach given in \cite{zou2012unified}, 
\begin{lemma}\label{moti}
Assume gain process follows, \ref{eq1} and the value of American option at time $t$, which mature at time $T$ is given by \ref{eqvalue}. Then,
\[V_t(T)=G_t+\int_{t}^{T}E[e^{-r(u-t)}(\mu_u-rG_u)\mathbbm{1}_{\{\,\tau^*\ge u\}\,}|\mathcal{F}_t]du\]
\end{lemma}
\begin{proof}
We will use following identity in the value function.
\begin{align}\label{eq6}
e^{-r\tau^*}G_{\tau^*}-e^{-rt}G_{t}&=\int_{t}^{\tau^*}d(e^{-ru}G_u)\nonumber\\
&=\int_{t}^{\tau^*}(-re^{-ru}G_udu+e^{-ru}dG_u)\nonumber\\
&=\int_{t}^{\tau^*}e^{-ru}(-rG_udu+\mu_udu+\sigma_udW_u)\nonumber\\
&=\int_{t}^{\tau^*}e^{-ru}((\mu_u-rG_u)du+\sigma_udW_u)\nonumber\\
e^{-r\tau^*}G_{\tau^*}&=e^{-rt}G_{t}+\int_{t}^{\tau^*}e^{-ru}((\mu_u-rG_u)du+\sigma_udW_u)
\end{align}
The value function $V_t(T)$ is given as
\begin{align*}
V_t(T) &=E[e^{-r(\tau^*-t)}G_{\tau^*} | \mathcal{F}_t]\\
&=e^{rt}E[e^{-r\tau^*}G_{\tau^*}|\mathcal{F}_t]\\
\text{using the identity given in \cref{eq6}}\\
&=G_t+E[\int_{t}^{\tau^*}e^{-r(u-t)}(\mu_u-rG_u)du|\mathcal{F}_t]\\ &+E[\int_{t}^{\tau^*}e^{-r(u-t)}\sigma_udW_u|\mathcal{F}_t]
\end{align*}
But $\int_{t}^{\tau^*}e^{-r(u-t)}\sigma_udW_u$ is a martingale. So, $E[\int_{t}^{\tau^*}e^{-r(u-t)}\sigma_udW_u|\mathcal{F}_t]=0$. This implies,
\begin{align}
V_t(T)&=G_t+E[\int_{t}^{\tau^*}e^{-r(u-t)}(\mu_u-rG_u)du|\mathcal{F}_t]\nonumber \\
&=G_t+E[\int_{t}^{T}e^{-r(u-t)}(\mu_u-rG_u)\mathbbm{1}_{\{\,\tau^*\ge u\}\,}du|\mathcal{F}_t]\nonumber \\
\text{By Fubini Theorem,}\nonumber\\
V_t(T)&=G_t+\int_{t}^{T}E[e^{-r(u-t)}(\mu_u-rG_u)\mathbbm{1}_{\{\,\tau^*\ge u\}\,}|\mathcal{F}_t]du \label{eq2}
\end{align}
\end{proof}
We recognize the integral is forward looking. Therefore motivated by lemma \ref{moti}, we define additive forward rate model for American option pricing as follows:
\begin{equation}\label{eq4}
    V_t(T)=G_t+\int_t^T f_t(u)du
\end{equation}
Assume the dynamics of the forward rate process, $\{\,f_t(T)\}\,_{t\in[0,T]}$ is given by 
\begin{equation}\label{eq5}
d_tf_t(T)=\alpha_t(T)dt+\beta_t(T)dW_t
\end{equation}
where $\int_{t}^{T}|\alpha_t(u)|du<\infty$ P-a.s and $\int_{t}^{T} \beta_t^2(u)du<\infty$ P-a.s.\\
From \ref{eq4}, and \ref{eq5} we can observe that,
\begin{equation}
\begin{aligned}
    d_tV_t(T) &=d_tG_t+d_t\bigg(\int_t^Tf_t(u)du\bigg)\\
    &=\bigg[\mu_t-f_t(t)+\int_t^T\alpha_t(u)du\bigg]dt+\bigg[\sigma_t+\int_t^T\beta_t(u)\bigg]dW_t
    \end{aligned}
\end{equation}

    Let's recall that under the HJM approach, a model must be be able to produce arbitrage free valuation. This can be achieve by imposing conditions on the drift of the forward rate. Following lemma guarantee the absence of arbitrage for the additive model.
\begin{lemma}
The additive model defined in \ref{eq4} is arbitrage free if \[ \alpha_t(T)=rf_t(T)\]
\end{lemma}
\begin{proof}
Consider the discounted value process, $e^{-rt}V_t(T)$, We can see that 
\begin{equation}
    \begin{aligned}
    d_t\big(e^{-rt}V_t(T)\big) &=e^{-rt}\bigg[-rV_t(T)+\mu_t-f_t(t)+\int_t^T\alpha_t(u)du\bigg]dt \\ &+e^{-rt}\bigg[\sigma_t+\int_t^T\beta_t(u)\bigg]dW_t
    \end{aligned}
\end{equation}
Since the discounted value process is a martingale, $$-rV_t(T)+\mu_t-f_t(t)+\int_t^T\alpha_t(u)du=0$$ By differentiating with respect to $T$, we get
    \begin{equation}
        \alpha_t(T)=r\pderiv{}{T}V_t(T)=rf_t(T)
    \end{equation}
    Hence the result.
\end{proof}

From \ref{eq4}, it is evident that, 
\begin{equation}
    f_t(T)=\pderiv{}{T} V_t(T) 
\end{equation}
Hence using \ref{eq2}, 
\begin{equation}\label{eqf}
f_t(T)=E[e^{-r(T-t)}(\mu_T-rG_T)1_{\{\,\tau^*\ge T\}\,}|\mathcal{F}_t]
\end{equation}

Let us introduce one of the integral part of HJM modeling philosophy known as the spot consistency condition. 

\begin{lemma} Define the spot consistency condition of additive forward rate model as $f_t(t)=\lim_{T\to t}f_t(T)$. Then $f_t(t)=\mu_t-rG_t$, where $\mu_t$ is the drift of the gain process given in \cref{eq1} and $G_t$ is the gain at time $t$, which is a known quantity.
\end{lemma}

\begin{proof}
Recall that $f_t(u)=E[e^{-r(u-t)}(\mu_u-rG_u)1_{\{\,\tau^*\ge u\}\,}|\mathcal{F}_t]$. Thus,
\begin{align*}
f_t(t)&=\lim_{T\to t}f_t(T)\\
&=\lim_{T\to t} E[e^{-r(T-t)}(\mu_T-rG_T)1_{\{\,\tau^*\ge T\}\,}|\mathcal{F}_t]\\
&\text{ but by dominated convergence theorem,}\\
&=E[\lim_{T\to t}e^{-r(T-t)}(\mu_T-rG_T)1_{\{\,\tau^*\ge T\}\,}|\mathcal{F}_t]\\
&=\mu_t-rG_t
\end{align*}
\end{proof}

So far we have proposed a new model to solve the optimal stopping problem, gave the conditions for the model to be arbitrage free and associated spot consistency condition. Now we will give the optimal stopping time criteria for the additive model. Recall that under the traditional approach, optimal stopping time is given in \cref{eq3}. The value function under the additive model is given in \cref{eq4}. By combining these two, we can obtain the optimal stopping time for the value process under the additive model as follows:

For the continuation region, we have $V_t(T)\ge G_t$ . Hence $G_t+\int_t^T f_t(u)du\ge G_t$. Thus, optimal stopping time is given when,
\begin{equation}\label{eq27}
\tau^*=\tau_t=\inf\{\,t\le s\le T|\int_{s}^{T}f_s(u)du=0 \}\,
\end{equation}

\subsection{Put option price under additive model}
In this section we demonstrate how to calculate American put option price under proposed additive forward $\{\,f_t(T)\}\,_{t\in [0,T]}$ modeling structure. With American put, $G_{t}=(K-S_t)^{+}$ and assume stock prices moves according to the Geometric Brownian motion with constant drift and constant volatility. Let the drift equal to risk free rate $r$ and constant volatility be given by $b$. This implies, $$\frac{dS_t}{S_t}=rdt+bdW_t$$ 
In order to get differentials of $dG_t$, we use Meyer-Tanaka formula \cite{karatzas2014brownian}. From Meyer-Tanaka formula, 
$(K-S_t)^+ =(K-S_0)^+ -\int_{0}^t \mathbbm{1}_{\{\,S_u<K\}\,}dS_u+\frac{1}{2}L_t^K(S_u)$ where $0\le u\le t$. Since American put has an early exercise boundary $l$, where $l\le K$, local time spent at level $K$, $L_t^K(S_u)$ is zero on the event $\{\,\tau^*\ge t\}\,$.

This implies,
\begin{equation}
\begin{aligned}
d(K-S_t)^{+} &= -d(\int_0^t \mathbbm{1}_{\{\,K>S_u\}\,}dS_u)\\
&= -\mathbbm{1}_{K>S_t} (rS_tdt+bS_tdW_t)\\
&= -rS_t \mathbbm{1}_{\{\,K>S_t\}\,}dt-bS_t\mathbbm{1}_{\{\,K>S_t\}\,} dW_t\\ \text{Hence,} & \\ 
dG_t &=-\big(rS_tdt+bS_tdW_t)\mathbbm{1}_{\{\,K>S_t\}\,}
\end{aligned}
\end{equation}
By comparing with the equation \ref{eq1}, we conclude $\mu_t=-rS_t\mathbbm{1}_{\{\,K>S_t\}\,}$ and $\sigma_t=-bS_t\mathbbm{1}_{\{\,K>S_t\}\,}$. Thus under the proposed additive forward rate model, the price of American put at time $t$ which expires at $T$ and exercise price of $K$ can be given as
\begin{equation}\label{eq31}
    V_t(T)=(K-S_t)^{+}+\int_{0}^Tf_t(u)du
\end{equation}
where $f_t(u)=E[e^{-r(u-t)}(\mu_u-rG_u)\mathbbm{1}_{\{\,\tau*\ge u\}\,}|F_t]$ where $t\le u\le T$. By substituting for $\mu_u$ and $G_u$ we can write $f_t(u)=E[e^{-r(u-t)}\big((-rS_u)-r(K-S_u)\big)\mathbbm{1}_{\{\,\tau*\ge u, K\ge S_u\}\,}|F_t]$. This expression can be simplified in to following form.
$$f_t(u)=-(rK)E[e^{-r(u-t)}\mathbbm{1}_{\{\,\tau*\ge u, K\ge S_u\}\,}|F_t]$$
By substituting in the equation \ref{eq31} we get
\begin{align*}
    V_t &=(K-S_t)^{+}-\int_{0}^T (rK)E[e^{-r(u-t)}\mathbbm{1}_{\{\,\tau*\ge u, K\ge S_u\}\,}|F_t]du \\
    &=(K-S_t)^{+}-rK \int_{t}^{\tau*}E[e^{-r(u-t)}\mathbbm{1}_{\{\, K\ge S_u\}\,}|F_t]du\\
    &=(K-S_t)^{+} -rK \int_t^{\tau*}e^{-r(u-t)}P(K>S_u|S_t)du\\
    &=(K-S_t)^{+}-rK\int_t^{\tau^*}e^{-r(u-t)}N(-d_2) du\\
\end{align*}
where $N(\cdot)$ denotes cumulative normal distribution function and $d_2=\frac{ln(S_t/K)+(r-\frac{b^2}{2})(u-t)}{b\sqrt{u-t}} $. This completes the example for American put under additive model.

\section{Multiplicative model}
Assume that the dynamics of the gain process $G_t$ is given by 
\begin{equation}\label{eq41}
dG_t=\mu_tG_tdt+\sigma_tG_tdW_t
\end{equation}
where $\int_{t}^{T}|\alpha_t(u)|du<\infty$ P-a.s and $\int_{t}^{T} \beta_t^2(u)du<\infty$ P-a.s. This imply \cite{privault2013stochastic} 
\begin{equation}\label{eq42}
G_t=G_0e^{\int_0^t (\mu_u-\frac{1}{2}\sigma^2_u)du+\int_0^t\sigma_udW_u}
\end{equation}

For American option price $V_t(T)$ and the gain process $G_t$, we define the 
value function of the American option under the multiplicative model as
\begin{equation}\label{eq44}
V_t(T)=G_te^{-\int_t^Tf_t(u)du}
\end{equation}
Then forward rate under the multiplicative model as
\begin{equation}\label{eq43}
  f_t(T)=-\pderiv{}{T}\ln V_t(T)  
\end{equation}

Assume the dynamics of the forward rate process, $\{\,f_t(u)\}\,_{t\in [0,T]}$ is given by 
\begin{equation}\label{eq45}
df_t(T)=\alpha_t(T)dt+\beta_t(T)dW_t
\end{equation}
where $\int_{t}^{T}|\alpha_t(u)|du<\infty$ P-a.s and $\int_{t}^{T} \beta_t^2(u)du<\infty$ P-a.s.\\
Under the multiplicative approach, spot consistency condition for HJM model is given by the following theorem.
\begin{lemma}\label{spotmult}
Define the spot consistency condition for forward drift model as $f_t(t)=\lim_{T\to t}f_t(T)$. Then $f_t(t)=r-\mu_t$ on $[0, \tau^*]$ where $r$ is the risk free interest rate.
\end{lemma}
\begin{proof}
From equation \ref{eq43}, \[f_t(T)=-\pderiv{}{T}\ln V_t(T) =\lim_{h\to 0} \frac{1}{h}\ln\frac{V_t(T)}{V_t(T+h)}\]
Thus, 
\[f_t(t)=\lim_{T\to t}f_t(T)=\lim_{T\to t}\lim_{h\to 0}\frac{1}{h}\ln\frac{V_t(T)}{V_t(T+h)}=\lim_{h\to 0}\ln \frac{V_t(t)}{V_t(t+h)}\]
But we know $V_t(t)=G_t$. Thus,
\[f_t(t)=\lim_{h\to 0}\frac{1}{h} \ln \frac{G_t}{V_t(t+h)}\]
Under the multiplicative model, when $t\le \tau^*\le t+h$
\begin{align*}
V_t(t+h)&=E[e^{-r(\tau^*-t)}G_{\tau^*}|F_t]\\
&=E[e^{-r(\tau^*-t)}G_t e^{\int_{t}^{\tau^*}(\mu_u-\frac{\sigma_u^2}{2})du+\int_{t}^{\tau^*}\sigma_udW_u}|F_t]\\
&= G_tE[ e^{\int_{t}^{\tau^*}(\mu_u-r-\frac{\sigma_u^2}{2})du+\int_{t}^{\tau^*}\sigma_udW_u}|F_t]\\
&= G_t e^{\int_{t}^{t+h}(\mu_u-r)\mathbbm{1}_{\{\,\tau^*\ge u\}\,}}du
\end{align*}
Thus,
\[f_t(t)=\lim_{h\to 0}\frac{1}{h}\int_{t}^{t+h} -(\mu_u-r)\mathbbm{1}_{\{\,\tau^*\ge u\}\,}du=(r-\mu_t)-\lim_{h\to 0}\frac{1}{h}\int_{t}^{\tau^*} (\mu_u-\mu_t) du\] As $u\to t$, we have $\mu_u\to \mu_t$. Hence by dominated convergence theorem, $\lim_{h\to 0}\frac{1}{h}\int_{t}^{\tau^*} |\mu_u-\mu_t| du\le \lim_{h\to 0}\frac{1}{h}\int_t^{\tau^*} \epsilon du$ for some $\epsilon>0$. Thus, 
$\lim_{h\to 0}\frac{1}{h}\int_{t}^{\tau^*} (\mu_u-\mu_t) du=0$ and \[f_t(t)=r-\mu_t\]
\end{proof}
No arbitrage condition for multiplicative model is given by the  following lemma.

\begin{lemma}\label{multidrift}
Under the multiplicative model we defined above,no arbitrage condition is given by $$\alpha_t(T)= \beta_t(T)\bigg(\int_{t}^T\beta_t(u)du-\sigma_t\bigg)$$ on $[0,\tau^*]$
\end{lemma}
\begin{proof}
 Define, $A(t)=\int_t^Tf_t(u)du$. Then we can write differentials of $V_t(T)$ as follows;
\begin{equation}
\begin{aligned}
    d_t(V_t(T))&=d_t(G_te^{-A(t)})\\
    &=G_td_t(e^{-A(t)})+e^{-A(t)}d_t(G_t)+d_t(G_t)d_t(e^{-A(t)})\\
    &=\bigg[G_te^{-A(t)}\bigg(f_t(t)dt-\big(\int_{t}^T\alpha_t(u)du\big)dt-\big(\int_{t}^T\beta_t(u)du\big)dW_t \\&-\frac{1}{2}\big(\int_{t}^T\beta_t(u)du\big)^2dt\bigg)\bigg]+ \bigg[G_te^{-A(t)}\bigg(\mu_tdt+\sigma_tdW_t\bigg)\bigg] \\
    &+\bigg[-G_te^{-A(t)}\sigma_t\bigg(\int_t^{T}\beta_t(u)du\bigg)dt\bigg] \\
    &=G_te^{-A(t)}\bigg[\bigg[f_t(t)-\big(\int_{t}^T\alpha_t(u)du\big)+\frac{1}{2}\big(\int_{t}^T\beta_t(u)du\big)^2+\mu_t \\&-\sigma_t\bigg(\int_t^{T}\beta_t(u)du\bigg)\bigg]dt+\bigg[-\big(\int_{t}^T\beta_t(u)du\big)+\sigma_t\bigg]dW_t\bigg]\\
  \end{aligned}
\end{equation}  
Hence,
\begin{equation}
\begin{aligned}\label{eqdvmult}
    \frac{d_t(V_t(T))}{V_t(T)}&=\bigg[f_t(t)-\big(\int_{t}^T\alpha_t(u)du\big)+\frac{1}{2}\big(\int_{t}^T\beta_t(u)du\big)^2+\mu_t \\ &-\sigma_t\bigg(\int_t^{T}\beta_t(u)du\bigg)\bigg]dt+\bigg[-\big(\int_{t}^T\beta_t(u)du\big)+\sigma_t\bigg]dW_t
\end{aligned}
\end{equation}
In order for model to be arbitrage free, the discounted value process must be a martingale. Hence, consider the discounted value process given by  $\{\,e^{-rt}V_t(T)\}_{t\ge 0},$. Now observe,
\begin{align*}
    d_t(e^{-rt}V_t) &=e^{-rt}V_t(T)\bigg[-rdt+d_t(V_t(T))\bigg]\\
    &=\bigg[-r+f_t(t)-\big(\int_{t}^T\alpha_t(u)du\big)+\frac{1}{2}\big(\int_{t}^T\beta_t(u)du\big)^2+\mu_t \\&-\sigma_t\bigg(\int_t^{T}\beta_t(u)du\bigg)\bigg]dt+\bigg[-\big(\int_{t}^T\beta_t(u)du\big)+\sigma_t\bigg]dW_t
\end{align*}
Observe that from spot consistency condition given in lemma \ref{spotmult}, $f_t(t)=r-\mu_t$ and the process becomes a martingale if,
\begin{align*}
-\big(\int_{t}^T\alpha_t(u)du\big)+\frac{1}{2}\big(\int_{t}^T\beta_t(u)du\big)^2-\sigma_t\bigg(\int_t^{T}\beta_t(u)du\bigg)&=0
\end{align*}
 By taking differentials with respect to $T$ we conclude that,
\[\alpha_t(T)=\beta_t(T)\bigg(\int_{t}^T\beta_t(u)du-\sigma_t\bigg)\]

\end{proof}
\begin{lemma}\label{multprice}
Assume the dynamics of the gain process follows equation \ref{eq41}, forward rates defined under multiplicative model given in equation \ref{eq43}-\ref{eq45}, then the value of the American option is given by, 
\begin{equation}
    \begin{aligned}
    V_t(T)= V_0(T)e^{\int_{0}^t \big(\gamma(s)-\frac{1}{2}\theta(s)^2\big)ds+\theta(s)dW_s}
    \end{aligned}
\end{equation}
where $\gamma(s)=r-\big(\int_{s}^T\alpha_s(u)du\big)+\frac{1}{2}\big(\int_{s}^T\beta_s(u)du\big)^2-\sigma_s\bigg(\int_s^{T}\beta_s(u)du\bigg)$
and $\theta(s)=-\big(\int_{s}^T\beta_s(u)du\big)+\sigma_s$
\end{lemma}
\begin{proof}
From equation \ref{eqdvmult}, and spot consistency condition in lemma \ref{spotmult}, this follows.
\end{proof}

Stopping time associated with multiplicative model is given by equation similar to \ref{eq27}. In other words, the optimal stopping time of the value process under multiplicative models is given by
\[\tau^*=\inf\{\,s\in[t,T]|:\int_{s}^Tf_s(u)du=0\}\,\]

\section{Multiplicative model example}
Consider a derivative with payoff given by $S_t^a$ where $a>0$. In this section we wish show how to calculate the price of this derivative using proposed multiplicative forward modeling structure. Suppose the underlying stock price follows geometric Brownian motion with a constant drift $r$ (risk-free rate) and constant volatility $b$. This implies,
\begin{equation}
\frac{dS_t}{S_t}=rdt+bdW_t
\end{equation}
Then payoff (gain) process $G_t=S_t^a$ have following dynamics
\begin{equation}
    \frac{dG_t}{G_t}=\bigg(ar+\frac{1}{2}a(a-1)b^2\bigg)dt+abdW_t
\end{equation}
Assume dynamics of the forward rate process is given by equation \ref{eq45}. 
Further assume, forward rate model is Gaussian  where $\beta_t(u)=\beta$ is constant. Then from lemma \ref{multidrift}, $\alpha_t(u)=\beta \bigg(\beta(u-t)-ab\bigg)$. Now using \ref{multprice}, the price of the derivative at time $t$ under proposed forward rate modelling method is 
\begin{equation}
    \begin{aligned}
    V_t(T)= V_0(T)e^{\int_{0}^t \big(\gamma(s)-\frac{1}{2}\theta(s)^2\big)ds+\theta(s)dW_s}
    \end{aligned}
\end{equation}
where $\gamma(s)=r$
and $\theta(s)=-\beta(T-s)+\sigma_s$. Observe that $V_0(T)$ is available in the market at time zero.

\section{Conclusion}
In this paper we have proposed a new approach for pricing American options using HJM framework.This includes an alternative pricing function, corresponding stopping time and stopping criteria. In future, we seek to implement and test this model using market data.

\section{Declaration of competing interest}
The authors declare that they have no competing interests.

\bibliographystyle{agsm}
\bibliography{ex_article}

\end{document}